\documentstyle[12pt,epsf]{article}
\begin{document}
\voffset -3.0cm
\textwidth=16cm
\textheight=22cm
\begin{titlepage}
\pagestyle{empty}
\begin{flushright}
\vbox{
{\bf TTP 94-4}\\
{\rm March 1994}
}
\end{flushright}
\renewcommand{\thefootnote}{\fnsymbol{footnote}}
\vspace{0.5cm}
\begin{center}
{\large V-A TESTS THROUGH LEPTONS
FROM POLARISED\\
TOP QUARKS
\footnote{ Work supported in part by
BMFT under contract 056KA93P and by KBN
under grant 2P30225206.}  }
\end{center}
\vskip1.0cm
\hyphenation{Germany}
\begin{center}
{\sc M. Je\.zabek}
\\
\vskip0.2cm
Institute of Nuclear Physics, Kawiory 26a, PL-30055 Cracow,
Poland \footnote{Permanent address, e-mail:jezabek@chopin.ifj.edu.pl}
\\
and\\
Institut f\"ur Theoretische Teilchenphysik, Universit\"at Karlsruhe\\
D-76128 Karlsruhe, Germany
\vskip0.7cm
{\sc J.H. K\"uhn}\\
\vskip0.2cm
Institut f\"ur Theoretische Teilchenphysik, Universit\"at Karlsruhe\\
D-76128 Karlsruhe, Germany
\end{center}
\vskip1.5cm
\begin{center}
Abstract
\end{center}
{\small Angular-energy distributions are studied for charged leptons
and neutrinos from the decays of polarised top quarks. A small
admixture of V+A interactions is incorporated. The polarisation
dependent part of the neutrino distribution which can be measured
experimentally through the missing momentum is particularly
sensitive towards deviations from the V-A structure. This result
remains unaffected by QCD corrections which, however, cannot be
neglected in a quantitative analysis.
\vskip1.0cm    }

\end{titlepage}
\setcounter{footnote}{0}
\renewcommand{\thefootnote}{\arabic{footnote}}
The analysis of polarised quarks and their decays, in particular
of top quarks, has recently attracted considerable attention.
Studies at a linear electron-positron collider are particularly
clean for precision tests \cite{Kuehn1,Kuehn2}. However, also
hadronic \cite{DPRK,Kuehn4,KLY}
or $\gamma\gamma$ collisions \cite{FKK}
and subsequent spin analysis of top quarks might reveal new
information. These studies will allow to determine the top
quark coupling to the $W$ and $Z$ boson, confirming the predictions
of the Standard Model or providing clues for physics beyond.
For the top quark this latter possibility is particularly intriguing
since it plays an exceptional role in the fermion mass spectrum.
In this article we point out that neutrino momentum distributions
are particularly sensitive towards violations of the V-A structure
of the charged current.
To perform such tests a sample of maximally polarised top quarks
is required. \\
A number of mechanisms have been suggested that will
lead to polarised top quarks.
For $\gamma\gamma$ collisions with circular polarised photons
this possibility has been discussed in \cite{FKK}. Related studies
may be performed in hadronic collisions which in this case, however,
are based on the correlation between $t$ and $\bar t$ decay products
\cite{Kuehn4,KRZ,KLY}. The most efficient and flexible
reactions to produce
polarised top quarks are electron-positron collisions. A small
component of polarisation transverse to the production plane
is induced by final state interactions which have been
calculated in perturbative QCD \cite{DPRK,KLY}.
The longitudinal polarisation $P_L$
is large. Its dependence on the production angle, beam energy
and the top mass is discussed in \cite{KRZ,AS1}. $P_L$ varies
strongly with the production angle, e.g. between nearly $0.6$ for
$\cos\vartheta\,=\,-1$ and $-0.3$ for $\cos\vartheta\,=\,1$
at $\sqrt{s}\,=\,500$ GeV. Averaging over the production angle
leads therefore to a significant reduction of $P_L$ with typical
values of $\langle P_L\rangle$ around 0.2 \cite{KPT}. QCD corrections
change $\langle P_L\rangle$ by a relative amount
of about 3\% \cite{KPT}. \\
All these reactions lead to sizable polarisation and can be used
to obtain information on the production mechanism. The most
efficient analyser of top polarisation is the charged lepton
direction (as seen from the top restframe) in semileptonic top
decays $t\to Wb\to \ell^+\nu b$. The decay distribution
factorises into an energy and an angular dependent part
\cite{KS,JK2}
\begin{equation}
{ {\rm d}N\over{\rm d}x_\ell\,{\rm d}\cos\theta_+}
\,=\, { {\rm d}N\over{\rm d}x_\ell}\,
\left( 1+ \cos\theta_+\right)/2
\end{equation}
where $\theta_+$ denotes the angle between the top spin and
the lepton direction.
This factorised form applies for arbitrary top mass below
and above the threshold for decays into real $W$ bosons.
QCD corrections to this distribution
are well under control \cite{CJK}
and will be included in the discussion below.
The analysing power of the charged lepton direction is obviously
maximal and hence far superior to the $W$ direction \cite{PSch}
which suffers from the suppression factor
$(m_t^2 - 2 m_{\rm w}^2)/(m_t^2 + 2 m_{\rm w}^2)$.
However, two drawbacks of all these proposals are evident:
production and decay are mixed in an intricate manner, and
furthermore the degree of polarisation is relatively small
and depends on the production angle. Top quark production
with longitudinally polarised electron beams and close to
threshold provides one important exception: the restricted
phase space leads to an amplitude which is dominantly S-wave
such that the electron (and positron) spin is directly
transferred to the top quark. Close to threshold and with
longitudinally polarised electrons one may deal with
a highly polarised sample of top quarks {\em independent
of the production dynamics}. Thus one may study the V-A
structure of $t$ decays under particularly convenient conditions:
large event rates, well identified restframe of the top quark,
and large degree of polarisation.\\
As stated before the angular distribution of charged leptons
is optimal for the analysis of the top polarisation. On the
other hand exactly for this reason it is less suited to identify
small admixtures of non-standard couplings, e.g. a small V+A
amplitude. In the analysis of neutrino ($\equiv$ missing
momentum)
distributions the situation is reversed: neutrino
distributions in top decays are sensitive to small V+A
admixtures and less sensitive to top polarisation.
This observation and the detailed analysis of charged
lepton and neutrino distributions
with a small admixture of V+A interaction
is the key point of our paper.
Two aspects will be considered. The strong sensitivity of
the neutrino energy-angular distribution towards a V+A admixture
will be demonstrated. In addition it will be shown that QCD
corrections to the Standard Model distributions are well
understood, such that any deviation can truly be attributed to a
deviation from pure V-A coupling in the decay.

As stated before, the emphasis of this work will be on a discussion
of polarised top quarks decays in their rest frame.
Nevertheless, a brief comment is appropriate on relativistic
top quarks produced far above threshold at a future linear collider.
With increasing
energy it will be more and more difficult to reconstruct the double
differential energy-angular distributions for the leptons.
For ultra-relativistic top quarks the situation will resemble
the one for $b$ quark semiloptonic decays at LEP energies.
A difference which makes the studies at the $Z^0$
peak particularly interesting is the very high polarisation
of $b$ quarks,
$P_L=\langle -0.94\rangle$,
which weakly depends on the production angle
\cite{Kuehn1}.
For top quarks the net polarisation will be smaller, as stated
earlier, but evidently a lot may be gained when considering
the asymmetry or the first Legendre moment
rather than the average over the production angle.
Moreover, the short lifetime in the top case helps to avoid
depolarisation in hadronization which is a serious problem for the
bottom quark. Thus the studies for ultra-relativistic top quarks are
also worth consideration. In close analogy to $b$ quarks at LEP
one can study the charged lepton
energy distribution (and its moments)
in the laboratory frame. Formulae are given in \cite{Mele92}
which relate this distribution to the double differential
angular-energy distribution discussed in the present article.
The neutrino distribution, which can be treated in the same way,
has recently attracted considerable interest.
It has been pointed out \cite{CJKK} that for
$b$ decays the neutrino distribution is highly sensitive to the
polarisation of the parent quark and can be used in polarisation
studies for $\Lambda_b$ baryons in addition to the distribution
of the charged lepton. In \cite{BR} it has been advocated that
the ratio of the average energies of the charged lepton
and the neutrino is particularly sensitive to $b$ polarisation.
In a recent preprint \cite{DW} a study has been proposed
of the charged weak current and its space-time structure
in $b$ semileptonic decays through the neutrino distribution.
Unlike for the top decay where the invariant mass of the lepton system
is equal to $m_{\rm w}$, the mass of $W$ boson, for $b$ decays
one usually integrates over this variable. We think, however, that
another difference is much more important. Neglecting the effects
of $W$ propagator the neutrino spectrum in the decays of a heavy
quark with the weak isospin $I_3= -1/2$ is the same as the
spectrum of the charged lepton for $I_3= 1/2$. Thus the results of
the present article suggest that for polarised $b$ quarks
the distribution of the charged lepton should be more sensitive
to V+A admixture. The detailed discussion will be given elsewhere.

Let us perform now a quantitative discussion  of
the sensitivity of different distributions to V+A admixtures.
The $tbW$ vertex is parametrised as follows:
$$ g_V\,\gamma^\mu\; +\; g_A\,\gamma^\mu\gamma_5$$
with
\begin{equation}
g_V = ( 1+ \kappa)/\sqrt{1+\kappa^2}\; , \qquad\qquad
g_A = ( -1+ \kappa)/\sqrt{1+\kappa^2}
\end{equation}
Hence $\kappa=0$ corresponds to pure V-A and $\kappa=\infty$
to V+A. We consider only leptonic decays of $W$.
The masses of all fermions in the final state are neglected.
The error associated with the lepton masses
is negligible at least for electrons and muons. The effects of
non-vanishing $b$ quark mass are of order $\kappa\epsilon$,
where $\epsilon= m_b/m_t$, so they are $\sim 1$\% for $\kappa=0.3$
and $\epsilon=0.03$. The following discussion refers to the rest
frame of the decaying top quark.\\
Neglecting QCD corrections the differential decay rate is
proportional to the following expression:
\begin{equation}
{\rm d}\Gamma \sim {1\over 1+\kappa^2}\,\left(\,
b\cdot\nu\; R_-\cdot\ell\; +\;\kappa^2\, b\cdot\ell\; R_+\cdot\nu\,
\right)
\end{equation}
where
$$ R_\pm\,=\,t \pm s$$
$t$, $b$, $\ell$ and $\nu$ are the scaled four-momenta of top,
bottom, charged lepton and neutrino. The common scaling factor
is $m_t^{-1}$, so
$$ t^2 = 1 \qquad {\rm and} \qquad
y = (\ell+\nu)^2 = (m_{\rm w}/m_t)^2\ .$$
$s = (0,\vec s)$ is the spin four-vector of the decaying quark.
$S= |\vec s| = 1$ corresponds to fully polarised and $S=0$ to
unpolarised top quarks. The following variables will be used:
\begin{equation}
\begin{tabular}{cc}
$x_\ell \; =\; 2\,t\cdot\ell$        & $\qquad\qquad
S\cos\theta_+ \; =\; -\, {s\cdot\ell / t\cdot\ell}$  \\
$x_\nu  \; =\; 2\,t\cdot\nu $        & $\qquad\qquad
S\cos\theta_0 \; =\; -\, {s\cdot\nu / t\cdot\nu}$         \\
\end{tabular}
\end{equation}
Thus, $x_\ell$ and $x_\nu$ are proportional to the energies
of the respective leptons, $\theta_+$ is the angle between $\vec s$
and the three-momentum of the charged lepton, and $\theta_0$
is the analogous angle for the neutrino. Employing the Dalitz
parametrization of the phase space and following the steps described
in Appendix B of \cite{CJK} and Appendix A.2 of \cite{CJ}
one derives the following formulae for the normalized double
differential angular and energy distributions for the charged
lepton and for the neutrino
\begin{eqnarray}
{ {\rm d}N\over{\rm d}x_\ell\,{\rm d}\cos\theta_+} &=&
{ 6\over\left(1+\kappa^2\right)\,{\cal F}_0(y)}
\left\{\, {\rm F}_0^+(x_\ell,y)\,+\, S\cos\theta_+\,
{\rm J}^+_0(x_\ell,y)  \right.
\nonumber\\
&& +\,\kappa^2\, \left. \left[\,
{\rm F}^-_0(x_\ell,y)\, -\, S\cos\theta_+\, {\rm J}^-_0(x_\ell,y)\,
\right]\, \right\}
\nonumber\\
&&\qquad\qquad\qquad\qquad\qquad\qquad\qquad\qquad
(\, y\le x_\ell \le 1\, )
\label{eq:elec}
\end{eqnarray}
\begin{eqnarray}
{ {\rm d}N\over{\rm d}x_\nu\,{\rm d}\cos\theta_0} &=&
{ 6\over\left(1+\kappa^2\right)\,{\cal F}_0(y)}
\left\{\, {\rm F}_0^-(x_\nu,y)\,+\, S\cos\theta_0\,
{\rm J}^-_0(x_\nu,y)  \right.
\nonumber\\
&& +\,\kappa^2\, \left. \left[ \,
{\rm F}^+_0(x_\nu,y)\, -\, S\cos\theta_0\, {\rm J}^+_0(x_\nu,y)\,
\right]\, \right\}
\nonumber\\
&&\qquad\qquad\qquad\qquad\qquad\qquad\qquad\qquad
(\, y\le x_\nu \le 1\, )
\label{eq:neut}
\end{eqnarray}
where
\begin{eqnarray}
{\rm F}^+_0(x,y) &=& x (1-x)
\label{eq:F0P}\\
{\rm J}^+_0(x,y) &=& {\rm F}^+_0(x,y)
\label{eq:J0P}\\
{\rm F}^-_0(x,y) &=& (x-y) (1-x+y)
\label{eq:F0M}\\
{\rm J}^-_0(x,y) &=& (x-y) (1-x+y-2y/x)
\label{eq:J0M} \\
{\cal F}_0(y) &=& 2 (1-y)^2 (1+2y)
\label{eq:F0Y}
\end{eqnarray}
These distributions can be easily integrated over the energies
leading to the following angular distributions
\begin{eqnarray}
{ {\rm d}N\over{\rm d}\cos\theta_+} &=&
{1\over 2}\,+\,
{1\over 2}\,S\cos\theta_+\,\left[ \,1-{\kappa^2\over 1+\kappa^2}\,
h(y)\, \right]
\label{eq:elec1}\\
{ {\rm d}N\over{\rm d}\cos\theta_0} &=&
{1\over2}\, -\, S\cos\theta_0\,
\left[\, 1\,-\, h(y)\,\right]\,\left[\, 1\,+\,
{\kappa^2 \over 1+\kappa^2}\, {h(y)\over 1-h(y)}\,
\right]
\label{eq:neut1}
\end{eqnarray}
where
\begin{equation}
h(y) = 2\,-\,{12y(1-y+y\ln y)\over (1-y)^2(1+2y)}
\end{equation}
For $m_{\rm w}=80$ GeV and $m_t=160$ GeV one obtains $y=0.25$ and
$h(0.25)=0.566$. It is
evident that the neutrino angular distribution is significantly
more sensitive towards the admixture of V+A than
the angular distribution of the charged lepton.
One can also gain sensitivity by studying
the double differential angular-energy distributions.
A convenient method is to calculate moments
of the distributions. The average energies
provide the simplest example:
\begin{eqnarray}
{ {\rm d}\langle x_\ell\rangle\over{\rm d}\cos\theta_+} &=&
{1+2y+3y^2\over 4(1+2y)}\,\left\{ \, \left[\,
1\,+\, {\kappa^2 \over 1+\kappa^2}\,{2y(1-y)\over 1+2y+3y^2}\,
\right]\right.
\nonumber\\
&& \left. +\, S\cos\theta_+\,\left[ \,1-{\kappa^2\over 1+\kappa^2}\,
{2(1-3y+2y^2)\over 1+2y+3y^2}\,
\right]\right\}
\label{eq:elec2}\\
{ {\rm d}\langle x_\nu\rangle\over{\rm d}\cos\theta_0} &=&
{1+4y+y^2\over 4(1+2y)}\,\left[\,
1\,-\, {\kappa^2 \over 1+\kappa^2}\,{2y(1-y)\over 1+4y+y^2}\,
\right]
\nonumber\\
&& +\, S\cos\theta_0\,  {1-8y+y^2\over 4(1+2y)}\,
\left[ \,1-{\kappa^2\over 1+\kappa^2}\,
{2(1-3y+2y^2)\over 1-8y+y^2}\,
\right]
\label{eq:neut2}
\end{eqnarray}
Hence the V-A form of the charged current in top decays
can be tested by measuring these averages. Even without
polarisation they are changed by an
admixture of V+A, so, in principle at least, it is possible
to see some effects although the `signal/background' ratio
is smaller. The best strategy, however, is to combine the
results on both the charged lepton and the neutrino distributions
for highly polarised top quarks.
At this point it becomes necessary to include QCD corrections
to the decay.

QCD corrections to the lepton spectra in decays of polarised quarks
were calculated previously
\footnote{The results of \cite{JK2} are in conflict with earlier
analytic calculations\cite{CCM}. The QCD correction to the lifetime
of $t$ quark \cite{JK1} follows from \cite{JK2} and has been
confirmed by other groups \cite{others}. The result of \cite{JK2}
on the $e^+$ spectrum in charm decays has been also confirmed
\cite{GWF}. For $e^-$ from $b$ decays see \cite{JK2},
footnote on page 27.
Adding this to the cross-checks performed in \cite{JK2,CJ}
the controversy can be considered as solved in favor of
our article \cite{JK2} and the related subsequent articles.}
\cite{JK2,CJK,CJKK,CJ}.
It is plausible that they reduce the V+A
contribution to the total rate by the same factor as the well-known
one for V-A contribution \cite{JK1}.
This implies that the ratio of V-A and V+A contributions
should be $1\, :\, \kappa^2$ also after inclusion of
${\cal O}(\alpha_s)$
corrections. This condition will be imposed on the QCD
corrected normalized distributions given in the following.
Anyway, the unknown corrections are small ($\sim \alpha_s \kappa^2$)
and the errors related to our procedure should be even smaller.
The normalized distributions including first order QCD corrections
read:
\begin{eqnarray}
{ {\rm d}N\over{\rm d}x_\ell\,{\rm d}\cos\theta_+} &=&
{1\over 2}\,\left[\,
{\rm A}_\ell(x_\ell)\, +\, S\cos\theta_+\,{\rm B}_\ell(x_\ell)\,
\right] \\
{ {\rm d}N\over{\rm d}x_\nu\,{\rm d}\cos\theta_0} &=&
{1\over 2}\,\left[\,
{\rm A}_\nu(x_\nu)\, +\, S\cos\theta_0\,{\rm B}_\nu(x_\nu)\,\right]
\end{eqnarray}
\pagebreak[4]
\begin{eqnarray}
{\rm A}_\ell(x) &=& {12\over 1+\kappa^2}\,
\left[\, { {\rm F}_0^+(x,y)\, -\,
a_s\, {\rm F}_1^+(x,y)  \over
{\cal F}_0(y)\, -\, a_s\, {\cal F}_1(y)}    \; +\;
{ \kappa^2\, {\rm F}^-_0(x,y) \over {\cal F}_0(y)}\, \right]
\label{eq:Aell}\\
{\rm B}_\ell(x) &=& {12\over 1+\kappa^2}\,
\left[\, { {\rm J}_0^+(x,y)\, -\,
a_s\, {\rm J}_1^+(x,y)  \over
{\cal F}_0(y)\, -\, a_s\, {\cal F}_1(y)}   \; -\;
{ \kappa^2\, {\rm J}^-_0(x,y) \over {\cal F}_0(y)}\;\, \right]
\label{eq:Bell}\\
{\rm A}_\nu(x) &=& {12\over 1+\kappa^2}\,
\left[\, { {\rm F}_0^-(x,y)\, -\,
a_s\, {\rm F}_1^-(x,y)  \over
{\cal F}_0(y)\, -\, a_s\, {\cal F}_1(y)}
\, +\, { \kappa^2\, {\rm F}^+_0(x,y) \over {\cal F}_0(y)}\; \right]
\label{eq:Anu}\\
{\rm B}_\nu(x) &=& {12\over 1+\kappa^2}\,
\left[\, { {\rm J}_0^-(x,y)\, -\,
a_s\, {\rm J}_1^-(x,y)  \over
{\cal F}_0(y)\, -\, a_s\, {\cal F}_1(y)}  \; -\;
{ \kappa^2\, {\rm J}^+_0(x,y) \over {\cal F}_0(y)}\;\, \right]
\label{eq:Bnu}
\end{eqnarray}
where
$$ a_s = {2\alpha_s\over 3\pi}\; ,$$
\begin{eqnarray}
{\cal F}_1(y) &=&
{\cal F}_0(y)\,
\left[\, {\textstyle {2\over3}}\pi^2+4{\rm Li}_2(y)
+2\ln y\ln(1-y)\,\right]
      - (1-y)(5+9y- 6y^2)
\nonumber\\  &&
      +\, 4y(1-y-2y^2)\ln y
      + 2(1-y)^2(5+4y)\ln(1-y)
\label{eq:F1y0}
\end{eqnarray}
as first derived in \cite{JK1},
and \cite{JK2,CJK,CJ}
\begin{eqnarray}
{\rm F}^+_1(x,y) &=&
   {\rm F}^+_0(x,y)\,\Phi_0 + x\Phi_1 - ( 3 + 2x +y )\Phi_{2\diamond 3}
   + 5(1-x)\Phi_4 + ( - 2xy
\nonumber\\
&&\   + 9x     - 4x^2 - 2y - y^2)\Phi_5
      + y  (  4 - 4x - y + y/x )/2
\label{eq:16}\\
\nonumber\\
{\rm F}^-_1(x,y) &=&
        {\rm F}^-_0(x,y)\,\Phi_0 + ( - 2xy  + x  + y  + y^2)\Phi_1
        + ( - 5  + 2x  - 3y )\Phi_{2\diamond 3}
\nonumber\\
&&
        + (  5 + 4xy - 5x + 3y - 5y^2 - 2y^2/x)\Phi_4
       + (  6xy + 9x  - 4x^2
\nonumber\\
&&\       - 11y   - 2y^2 + 2y^2/x )\Phi_5
        + y  (  2 + 3x - 3y - 2y/x )/2
\label{eq:17}\\
\nonumber\\
{\rm J}^+_1(x,y) &=&
       {\rm J}^+_0(x,y)\,\Phi_0 - x\Phi_1
       +  ( 5 - 2x - y - 2y/x - 2/x )\Phi_{2\diamond 3}
       +  ( - 3 + x
\nonumber\\
&&\        + 2/x)\Phi_4
       +  ( - 2xy + 3x - 4x^2 + 6y - y^2 - 2y^2/x)\Phi_5
\nonumber\\
&&       +  ( 2 - 2x^2 + 2y - 3y^2 - 2y/x + 3y^2/x )/2
\label{eq:18}\\
\nonumber\\
{\rm J}^-_1(x,y) &=&
      {\rm J}^-_0(x,y)\, \Phi_0
      + ( - 2xy - x  - 5y + y^2 -2y^2/x )\Phi_1
      + ( 3 + 10x + y
\nonumber\\
&&\
      + 10y/x - 2/x )\Phi_{2\diamond 3}
      + ( - 3 + 12xy + x - 7y - y^2 - 12y/x
\nonumber\\
&&\
+ 8y^2/x + 2/x )\Phi_4
+ ( 6xy - 9x - 4x^2 - y - 2y^2 + 10y^2/x)\Phi_5
\nonumber\\
&&         +  ( 2  - 5xy - 2x^2 + 2y + 7y^2 - 2y/x - 2y^2/x )/2
\label{eq:19}
\end{eqnarray}
where
\begin{eqnarray}
&\Phi_0\ \ &=\ {\textstyle {\pi^2\over 3}} + 2{\rm Li}_2(x)
 + 2{\rm Li}_2(y/x) + \ln^2\left({\textstyle {1-y/x\over 1-x}}\right)
\nonumber\\
&\Phi_1\ \ &=\  {\textstyle {\pi^2\over 6}} + {\rm Li}_2(y)
           - {\rm Li}_2(x)   - {\rm Li}_2(y/x)
\nonumber\\
&\Phi_{2\diamond 3} &=\ {\textstyle{1\over2}} (1-y)\ln(1-y)
\nonumber\\
&\Phi_4\ \ &=\ {\textstyle{1\over2}}\ln(1-x)
\nonumber\\
&\Phi_5\ \ &=\ {\textstyle{1\over2}}\ln(1-y/x)
\end{eqnarray}
It is worth mentioning that in \cite{CJ} formulae
are also given  for the QCD corrections to V-A heavy quark decays
for non-zero mass of the quark produced in the decay
which are only slightly more complicated
than those in the massless approximation. Thus it is easy to
generalize the present discussion to non-zero $\epsilon$. As stated
before, terms $\sim\kappa\epsilon$ appear for $\epsilon\ne 0$.
The corresponding QCD corrections of
${\cal O}(\alpha_s\kappa\epsilon)$
are unknown but small, and can be safely neglected. Anyhow, in the
present article we stick to the massless approximation for the
sake of simplicity of the presentation.

\begin{table}[h]
\begin{center}
\begin{tabular}{|c|r|r|r|r|}      \hline
   &          &          &          &     \\
$  k  $ &${\cal A}^{(\ell)}_k $ &${\cal B}^{(\ell)}_k$ &
${\cal A}^{(\nu)}_k$ & ${\cal B}^{(\nu)}_k$  \\
   &          &          &          &     \\
\hline
   -1   &    2.008  &    2.005  &   1.593  &  -0.707 \\
\cline{2-5}
        &     .981  &     .940  &   1.023  &   1.166 \\  \hline
    0   &    1.000  &     .998  &   1.000  &  -0.452 \\  \cline{2-5}
        &    1.000  &     .949  &   1.000  &   1.110 \\  \hline
    1   &     .559  &     .558  &    .683  &  -0.322 \\  \cline{2-5}
        &    1.021  &     .960  &    .984  &   1.068 \\  \hline
    2   &     .345  &     .344  &    .500  &  -0.249 \\  \cline{2-5}
        &    1.043  &     .973  &    .973  &   1.037 \\  \hline
    3   &     .230  &     .230  &    .385  &  -0.203 \\  \cline{2-5}
        &    1.064  &     .989  &    .965  &   1.015 \\  \hline
\end{tabular}
\end{center}
\caption{Moments of the coefficient functions defining the
angular-energy distributions for the charged leptons
and for the neutrinos
for $y=0.25$ and $\alpha_s= 0.11$:
the upper entries denote
the moments for $\kappa^2=0$
and the lower ones are the ratios between the moments for
$\kappa^2=0.1$ and $\kappa^2=0$.}
\end{table}
\noindent
In Table 1 the following moments are given
\begin{equation}
\begin{tabular}{cc}
${\cal A}^{(\ell)}_k \;=\; \int_y^1{\rm d}y\, x^k\,{\rm A}_\ell(x,y)$ ,&
$\qquad\qquad {\cal
B}^{(\ell)}_k = \int_y^1{\rm d}y\, x^k\,{\rm B}_\ell(x,y)$
\end{tabular}
\end{equation}
\begin{equation}
\begin{tabular}{cc}
${\cal A}^{(\nu)}_k  \;=\; \int_y^1{\rm d}y\, x^k\,{\rm A}_\nu(x,y)$ ,
  & $\qquad\qquad  {\cal
B}^{(\nu)}_k = \int_y^1{\rm d}y\, x^k\,{\rm B}_\nu(x,y)$
\end{tabular}
\end{equation}
\begin{figure}
\begin{center}
\leavevmode
\epsffile[70 320 550 500]{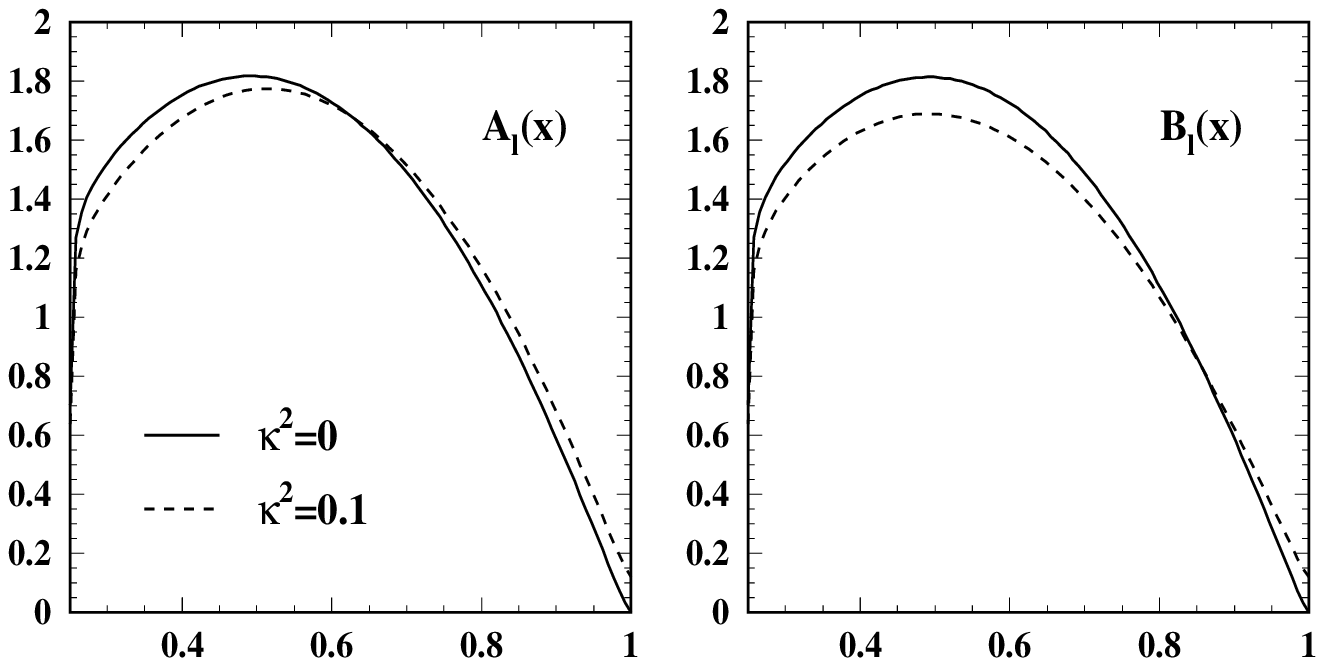}
\caption{The coefficient functions a) ${\rm A}_\ell(x)$ and
b) ${\rm B}_\ell(x)$ defining the charged lepton angular-energy
distribution for $y=0.25$ and $\alpha_s(m_t)=0.11$ : \ \
$\kappa^2=0$ -- solid lines and $\kappa^2=0.1$ -- dashed lines.
\label{fig-1} }
\end{center}
\begin{center}
\leavevmode
\epsffile[70 320 550 500]{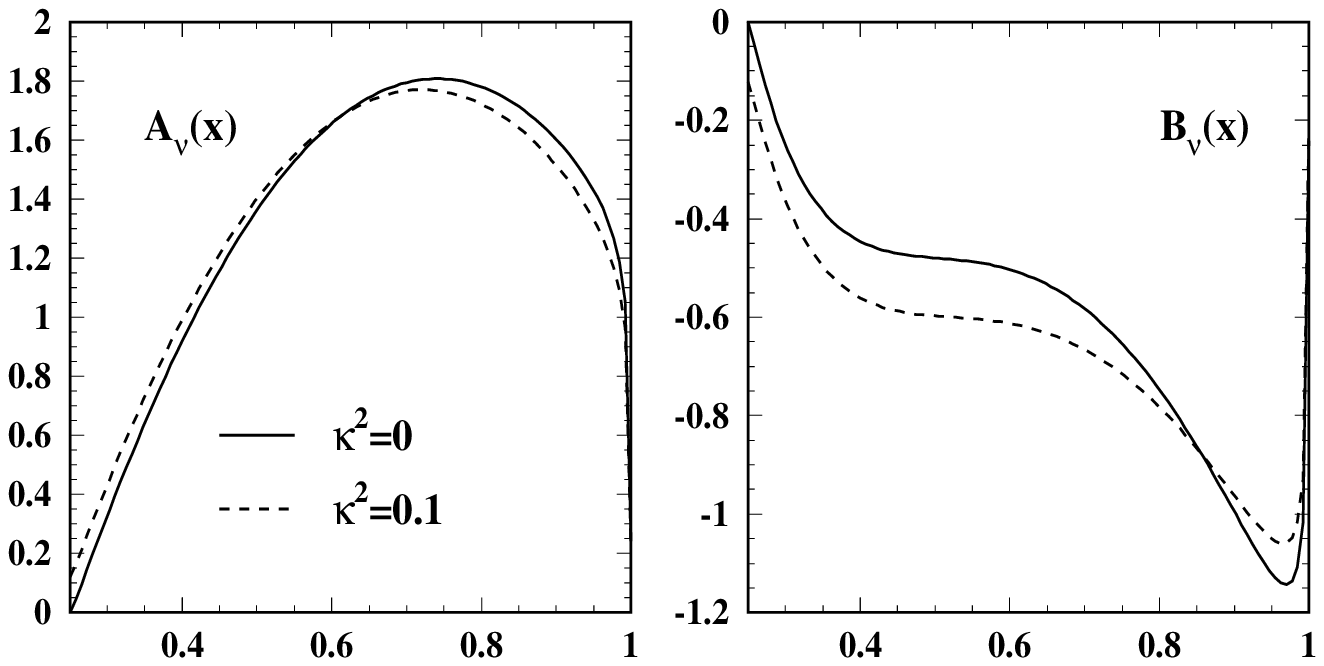}
\caption{The coefficient functions a) ${\rm A}_\nu(x)$ and
b) ${\rm B}_\nu(x)$ defining the neutrino angular-energy
distribution for $y=0.25$ and $\alpha_s(m_t)=0.11$ : \ \
$\kappa^2=0$ -- solid lines and $\kappa^2=0.1$ -- dashed lines.
\label{fig-2}  }
\end{center}
\end{figure}

\noindent
for integer $k$ between -1 and 3,
$y=0.25$ and $\alpha_s(m_t)= 0.11$~.
The upper entries in the table
denote the values of the moments for $\kappa^2=0$ and the
lower ones the ratios of the moments evaluated for
$\kappa^2=0.1$  to those  for $\kappa^2=0$.
From the comparison of the moments it is again evident that the
moments ${\cal B}_k^{(\nu)}$ which govern the angular dependence
of the neutrino spectrum
are particularly sensitive towards a V+A admixture. The effect
is most pronounced for the moment $k= -1$ which enhaces the lower
energy part of the spectrum and where the relative change amounts
to 17\% for $\kappa^2=0.1$~.\\
The same conclusions follow from Figs.\protect\ref{fig-1} and
\protect\ref{fig-2}
where the coefficient functions defined in eqs.
(\protect\ref{eq:Aell})-(\protect\ref{eq:Bnu}) are shown as
solid lines for $\kappa=0$ and as dashed lines for $\kappa=0.1$~.
In Figs.\ref{fig-1}a and \ref{fig-2}a the functions
${\rm A}_\ell(x)$ and ${\rm A}_\nu(x)$
are plotted for $y=0.25$, $\alpha_s(m_t)= 0.11$ and the functions
${\rm B}_\ell(x)$ and ${\rm B}_\nu(x)$ are shown in
Figs.\ref{fig-1}b and \ref{fig-2}b, respectively.

\begin{table}[h]
\begin{center}
\begin{tabular}{|l|l|c|c|c|c|}      \hline
  &   &          &          &          &     \\
$\alpha_s$& $\kappa^2$ & ${\rm A}_{\ell}(0.5) $ &
${\rm B}_{\ell}(0.5)$ & ${\rm A}_{\nu}(0.7)$&
${\rm B}_{\nu}(0.7)$  \\
  &   &          &          &          &     \\
\hline
0.11 & 0     &  1.816  &   1.812   &  1.794   &  -0.583   \\  \hline
0.11 & 0.1   &  1.772  &   1.688   &  1.767   &  -0.666   \\  \hline
0    & 0     &  1.778  &   1.778   &  1.760   &  -0.526   \\  \hline
0    & 0.1   &  1.737  &   1.657   &  1.736   &  -0.614   \\  \hline
\end{tabular}
\end{center}
\caption{Comparison of the coefficient functions with and without
QCD corrections and V+A admixture for $y=0.25$}
\end{table}
\noindent
The effect of QCD corrections is illustrated by
the comparison of the coefficient functions
${\rm A}_\ell(0.5)$,  ${\rm B}_\ell(0.5)$,
${\rm A}_\nu(0.7)$ and ${\rm B}_\nu(0.7)$ which are
given in Table 2. The effect of QCD
correction can mimic a small admicture of V+A interaction.
Therefore, inclusion of the radiative QCD correction to
the decay distributions is necessary for a quantitative study.

We conclude that the angular-energy distribution of neutrinos
from the polarised top quark decay will allow for a particularly
sensitive test of the V-A structure of the charged current.\\

\end{document}